\documentstyle[12pt,aasms4]{article}
\input psfig.sty

\begin{document}

\title{THE CRAB NEBULA: INTERPRETATION OF CHANDRA OBSERVATIONS}


\author{S.~V.Bogovalov,D.~V.~Khangoulyan}
\affil{Moscow state Engineering Physics Institute (Technical
University), Kashirskoe sh. 31, Moscow, 115409 Russia, E-mail:
bogoval@axpk40.mephi.ru }

\begin{abstract}
We interpret the observed X-ray morphology of the central part of
the Crab Nebula (torus~$+$~jets) in terms of the standard theory
by Kennel and Coroniti~(1984). The only new element is the
inclusion of anisotropy in the energy flux from the pulsar in the
theory. In the standard theory of relativistic winds, the Lorentz
factor of the particles in front of the shock that terminates the
pulsar relativistic wind depends on the polar angle as
$\gamma=\gamma_0+\gamma_m\sin^2\theta$, where $\gamma_0 \sim 200$
and $\gamma_m \sim 4.5\times 10^6$. The plasma flow in the wind is
isotropic. After the passage of the pulsar wind through the shock,
the flow becomes subsonic with a roughly constant (over the
plerion volume) pressure $P={1\over 3}n\epsilon$, where $n$ is the
plasma particle density and $\epsilon$ is the mean particle
energy. Since $\epsilon \sim \gamma mc^2$, a low-density region
filled with the most energetic electrons is formed near the
equator. A bright torus of synchrotron radiation develops here.
Jet-like regions are formed along the pulsar rotation axis, where
the particle density is almost four orders of magnitude higher
than that in the equatorial plane, because the particle energy
there is four orders of magnitude lower. The energy of these
particles is too low to produce detectable synchrotron radiation.
However, these quasi-jets become comparable in brightness to the
torus if additional particle acceleration takes place in the
plerion. We also present the results of our study of the
hydrodynamic interaction between an anisotropic wind and the
interstellar medium. We compare the calculated and observed
distributions of the volume intensity of X-ray radiation.

\vspace*{10pt} \emph{Key words:} plasma astrophysics,
hydrodynamics and shock waves.
\end{abstract}

\section*{INTRODUCTION}

The Crab Nebula is one of the most interesting and best-studied sources in the sky. This
object was observed over a wide wavelength range: from radio to gamma-rays with a photon
energy of 50~TeV (Aharonian~2000; Hester~1998; Shklovskii~1968). However, advances in
observational astronomy provide new data on the Crab Nebula. In the last decade,
progress in the technology of X-ray telescopes has allowed the Crab Nebula to be
observed with an angular resolution comparable to the angular resolution of ground-based
optical telescopes. As would be expected, this gave completely new information on the
structure of the central part of the nebula. The data obtained on the Chandra X-ray
observatory arouse particular interest.

The Chandra observations show that the central part of the nebula consists of two
components in the soft X-ray emission: a toroidal structure surrounding the pulsar
PSR~$0531+21$ and two jet-like features located perpendicular to the torus and emerging
from the pulsar (Weisskopf \emph{et al.}~2000). Interestingly, such a structure of the
plerion central region is also observed in the Vela pulsar (Pavlov \emph{et al.}~2001)
and in the supernova remnant G$0.9+1$ (Gaensler~2001). In this paper, we focus our
attention on the Crab Nebula primarily because of the parameters of the relativistic
plasma flow (pulsar wind) from this pulsar. Weak magnetation of the wind from
PSR~$0531+21$ allows the problem of its interaction with the interstellar medium to be
simplified to an extent that the intensityÐy distribution of synchrotron radiation in the
plerion can be easily estimated and compared with the observed one.

The first impression that arises when studying the Chandra images of the Crab Nebula is
that the toroidal structure surrounding the pulsar suggests the presence of an accretion
disk around the nebula. However, this interpretation of the observed picture is incorrect
for obvious reasons. First, there is no independent evidence for the existence of a
second companion and an accretion disk around the Crab pulsar. Second, the
characteristic size of the toroidal structure itself, $\sim10^{18}$~cm, rules out the
possibility of the disk interpretation of the observed picture.

The situation with the jet-like features is more complicated. Jets
are observed from many Galactic YSO  (Livio~1999), SS~433
(Cherepashchuk~1998), superluminal sources (Mirabel and
Rodrigues~1998) and AGN (Urri and Padovani~1995). A direct analogy
between the jets from these objects and those observed in the Crab
Nebula suggests itself. The assumption that the pulsar itself
ejects collimated plasma flows seems reasonable enough. However,
it is most likely incorrect.

The Crab Nebula is a typical plerion --- a bubble of relativistic
particles frozen in a magnetic field. This bubble is formed when
the flow of supersonic relativistic plasma (pulsar wind) ejected
by the pulsar interacts with the interstellar medium. Since the
wind itself is cold, it remains unobservable thus far (see,
however, Bogovalov and Aharonian~2000). After passing through the
wind-terminating shock, the particles are isotropized and begin to
emit synchrotron photons over a wide electromagnetic spectral
range, producing the observed plerion radiation; hence the
fundamental difference between the jet-like features observed in
the nebula and the actual jet flows observed in the Universe from
other objects. The latter are the supersonic collimated flows
ejected from the source. They are characterized by termination
when interacting with the interstellar medium to produce a shock
and the so-called lobes (Ferrari \emph{et al.} 1996). The jet-like
features themselves in the Crab Nebula are formed behind the shock
that terminates the pulsar wind. The plasma flow in them is
definitely subsonic. Therefore, the physics of this phenomenon
undoubtedly differs from the physics of the processes that give
rise to astrophysical jets.

Since the observed structures in the Crab Nebula result from the interaction of the
pulsar wind with the ambient medium, the pattern of this interaction must be studied to
understand their nature. The interaction of the wind from the Crab pulsar with the
interstellar medium has been analyzed by many authors. Rees and Gunn~(1974) and Kennel
and Coroniti~(1984) first gave important constraints on the parameters of the pulsar
wind immediately in front of the shock. The calculated plerion expansion velocity,
luminosity, and synchrotron radiation spectrum (from optical wavelengths to X-rays)
agree with the observed ones if the wind consists of electrons and positrons with a
Lorentz factor of $\sim 3\times 10^6$ and if almost all of the pulsar rotational losses
are transformed into the particle kinetic energy so that the ratio of the
electromagnetic energy flux to the particle kinetic energy flux is $\sigma = 3\times
10^{-3}$. For such wind parameters, we can also naturally explain the gamma-ray emission
from the Crab Nebula above 10~GeV, which is generated by the inverse Compton scattering
of the same electrons that generate the synchrotron radiation (Atoyan and
Aharonian~1996; de~Jager and Harding~1992).

The success of the theory by Kennel and Coroniti~(1984) was achieved through a
significant simplification of the problem. This theory assumes the problem to be
spherically symmetric. As long as the analysis was restricted to the integrated
characteristics of the radiation from the Crab Nebula (spectra, luminosity), this
limitation was not fundamental in nature. However, the observed X-ray morphology of the
Crab Nebula cannot be explained in terms of this theory. Clearly, the Crab Nebula is not
spherically symmetric. The more realistic pattern of interaction between an anisotropic
pulsar wind and the interstellar medium must be considered. Here, we made the first step
in solving this problem. At this stage, we do not set the goal of developing a
full-blown theory of the interaction between an anisotropic, magnetized pulsar wind and
a homogeneous interstellar medium. This is not yet possible. Here, we determine the
pattern of pulsar-wind anisotropy by using the results that have been obtained in pulsar
physics in recent years. We also perform a semiquantitative analysis of the result of
the interaction between such a wind and the interstellar medium, including an estimation
of the synchrotron radiation. In the end, we wish to understand whether the level of
anisotropy in pulsar winds that follows from the pulsar theory is enough to explain, at
least in general terms, the structure of the central part of the Crab Nebula observed on
the Chandra observatory.

\section*{THE PULSAR WIND FROM PSR~0531~+~21}

The integrated characteristics of the Crab Nebula can be naturally explained in terms of
the theory by Kennel and Coroniti~(1984) if the wind magnetization parameter is
$\sigma=3\times 10^{-3}$ and the wind Lorenz factor is $\gamma \sim 3\times 10^6$
immediately in front of the shock (Kennel and Coroniti~1984), although other wind
parameters cannot be completely ruled out either (Begelman~1998). The Kennel--Coroniti
theory naturally accounts for the source spectrum over an unprecedentedly wide
wavelength range, from optical to hard gamma-ray emission (fifteen orders of magnitude
in wavelength!). No theory in astrophysics can boast a similar success. However, there
is one problem in this theory that spoils the overall picture. It is not yet clear how
the pulsar PSR~$0531+21$ produces the relativistic wind with such parameters. This
remains one of the key puzzles in the physics of radio pulsars.

The problem is that all the currently available theories of particle acceleration and
plasma formation in pulsar magnetospheres [the polar-cap theory (Arons~1983; Daugherty
and Harding~1996) or the outer-gap theory (Romani~1996; Cheng \emph{et al.}~2000)] are
capable of explaining how the dense $e^{\pm}$~plasma that produces a relativistic
particle wind is formed. However, this plasma carries a negligible fraction ($\sim
10^{-4}$) of all rotational losses from the Crab pulsar. The entire energy flux from the
pulsar is concentrated in the electromagnetic field carried away by the wind, which
corresponds to the wind magnetization parameter $\sigma \sim 10^{4}$. The wind itself has
a modest Lorentz factor $\sim200$ near the radio-pulsar light cylinder (Daugherty and
Harding~1996). It is yet tbe clarified through which processes almost the entire
electromagnetic energy flux is transformed into the wind-particle kinetic energy on the
way from the light cylinder to the shock front, although substantial efforts were spared
to solve this problem (Coroniti~1990; Lyubarsky and Kirk~2001).

There is no need to know the wind acceleration mechanism to
determine the energy-flÐmux distribution in the wind. The fact that
the conservation of the energy flux in the wind holds in any case
is suffice. The point is that the electromagnetic energy flux in
the winds from radio pulsars propagates along streamlines. The
plasma kinetic energy flux also propagates along these
streamlines. This conclusion is based on MHD models of the winds
from axisymmetrically rotating objects (Okamoto~1978). However, it
has recently been shown to be also valid for obliquely rotating
objects where the flow beyond the light cylinder is concerned
(Bogovalov~1999). This ensures that the total energy flux
per~particle is conserved along a given streamline. The kinetic
energy flux per~particle, in units of~$mc^2$, is $\gamma$. The
electromagnetic energy flux per~particle, in units of~$mc^2$, is

\begin{equation}
 s={c\over 4\pi}[E\times B]/(n v mc^2\gamma),
\end{equation}
where $E$ is the electric field, $B$ is the magnetic field, $n$ is
the particle density in the intrinsic frame of reference, and $v$
is the wind velocity. The sum $\gamma + s$ depends on the
streamline but is conserved along it. It follows from the solution
of the problem on the structure of the wind from an oblique
rotator (Bogovalov~1999) that the energy flux in the wind
sufficiently far from the light cylinder may be considered to be
azimuthally symmetric, although the electromagnetic field itself
is not azimuthally symmetric in this case (see Bogovalov~(1999)
for details). In this notation, the magnetization parameter is
$\sigma = s/\gamma$. The conservation of the total energy flux
ensures that

\begin{equation}
\gamma_0+s_0=\gamma+s.
\end{equation}
The subscript~`0' marks the values near the light cylinder. Since $s \ll \gamma$ in
front of the shock that terminates the pulsar wind, the Lorentz factor of the preshock
plasma may be assumed to be $\gamma=\gamma_0+s_0$. Thus, the dependence of the particle
energy on the streamline along which the particles move is determined by their initial
energy and the initial distribution of the electromagnetic energy flux. We make the only
assumption. Assume that the initial distribution of the electromagnetic energy flux does
not depend on (or is almost independent of) the wind acceleration. This assumption holds
true in all cases if the wind is accelerated sufficiently far from the light cylinder in
the supersonic flow region. Then, the field distribution in the magnetosphere and, hence,
the initial electromagnetic energy flux do not depend on what happens downstream of the
magnetosonic surface. This is because no MHD signal can penetrate from the supersonic
flow region into the subsonic flow region and affect the flow in this region. If this is
the case, then it will suffice to determine~$s_0$, provided that there is no
acceleration.

Numerical and analytical calculations of the relativistic plasma flow show that the
plasma magnetic collimation is negligible for Lorentz factors $\gamma > 200$
(Beskin~1998; Bogovalov and Tsinganos~1999; Bogovalov~2001). The pulsar wind may be
assumed to spread out radially. Since the relation $E={r \sin\theta\Omega\over c} B_p$
(Mestel~1968), where $r$ is the distance to the pulsar, $\theta$ is the polar angle,
$\Omega$ is the pulsar angular velocity, and $B_p$ is the poloidal magnetic field in the
wind, holds between the electric field in the wind and the poloidal magnetic field, the
condition for the magnetic field being frozen in the plasma $E+{1\over c}[E\times B]=0$
takes the form

\begin{equation}
r \sin\theta\Omega B_p+B_{\varphi}v_p=v_{\varphi}B_p.
\end{equation}
At $r \gg c/\Omega$, the plasma angular velocity tends to zero,
because the angular momentum of the plasma particles
($r\sin(\theta)v_{\varphi}$) is limited above. We then derive a
simple expression for the toroidal field in the wind far from the
pulsar, $B_{\varphi}= {r \sin\theta\Omega\over v_p} B_p$.
Therefore, the electromagnetic energy flux per~particle (in units
$mc^2$) is

\begin{equation}
s_0=\sin^2\theta \left({\Omega r\over v_p}\right)^2{B_p^2\over
4\pi n_0mc^2\gamma_0}.
\end{equation}
We see that when the relativistic wind spreads out radially and uniformly, the Lorentz
factor~$\gamma_1$ of the preshock wind particles must have a latitude dependence of the
form

\begin{equation}
\gamma_1=\gamma_0+\gamma_m\sin^2\theta, \label{g1:Bogovalov_n}
\end{equation}
where $\gamma_m = ({\Omega r\over v_p})^2{B_p^2\over 4\pi n_0mc^2\gamma_0}$. This is the
maximum Lorentz factor of the preshock wind particles. To be consistent with the theory
by Kennel and Coroniti~(1984), it must be of the order of~$3 \times 10^6$.

Note that expression~(5) for the particle Lorentz factor
immediately follows from the MHD theory of magnetized winds from
rotating objects and is almost model-independent. It only assumes
that the particle flux from the pulsar is isotropic. Clearly, this
assumption does not severely restrict the range of applicability
of our results. For the standard parameters $\gamma_0=200$ and
$\gamma_m\sim 3\times 10^6$, the Lorentz factor changes with
latitude by four orders of magnitude. Even if the particle flux
changes with latitude by several times (or several tens of times),
this does not change the overall dependence. Anyway, the most
energetic particles will be near the equator and their Lorentz
factor will be higher than the Lorentz factor of the particles at
the rotation axis by several orders of magnitude.

Below, for our calculations, we assume the wind flow to be radial with an isotropic mass
flux. The plasma density in the intrinsic frame of reference in such a wind is

\begin{equation}
n_1={\dot N \over 4\pi v_p r^2\gamma}.
\end{equation}
Here, $\dot N$ is the rate of particle injection into the nebula. The toroidal magnetic
field has the distribution

\begin{equation}
B_{\varphi}= {B_0 r_0\sin\theta\over r},
\end{equation}
The quantity $B_0$ can be determined from the condition $\sigma=3\times 10^{-3}$. We
ignore the poloidal magnetic field, because it is much weaker than the toroidal magnetic
field ahead of the shock front. The plasma Lorentz factor is given by expression~(5).
Our objective is to determine (at least qualitatively) how a wind with the above
parameters interacts with a homogeneous ambient interstellar medium.

\section*{THE APPROXIMATIONS}

The problem of the interaction between a highly anisotropic, magnetized relativistic wind
and a homogeneous ambient medium has no analytic solution. Even obtaining a numerical
solution seems problematic so far. Below, we make two reasonable simplifications that
will allow us to answer the questions of interest by using simple mathematics.

(1) \emph{The approximation of a hydrodynamic interaction.} In the special case of the
Crab Nebula, as a first approximation, we may disregard the magnetic-field effect on the
postshock plasma dynamics. This approximation seems reasonable, because the pulsar wind
from PSR~$0531+21$ is weakly magnetized. Recall that the ratio of the preshock Poynting
flux to the plasma kinetic energy flux is $\sigma=3\times 10^{-3}$. Although the
magnetic field increases in strength by a factor of~3 after the shock passage, the ratio
of magnetic pressure to plasma pressure is small up to distances approximately equal to
five shock radii (see Kennel and Coroniti~1984). Further out, the magnetic pressure is
higher than the plasma pressure and it cannot be ignored. However, the region
of~$5r_{\mathrm{sh}}$ completely suits us. It is in this region that the most
interesting features of the central part of the Crab Nebula are formed: the X-ray torus
and the jet-like features. Therefore, below, the wind interaction is considered as a
purely hydrodynamic one. We will determine the magnetic-field evolution from the
induction equation with a given distribution of the plasma velocity field. Note that
although the magnetic-field effect on the plasma dynamics immediately behind the shock
wave is marginal, Lyubarsky~(2002) attempts to explain the observed jets in the Crab
Nebula as resulting from a magnetic compression of the wind after the shock wave. Below,
we show that these jets are formed in the hydrodynamic approximation without any
involvement of the magnetic field.

(2) \emph{The quasi-stationary approximation.} Strictly speaking, the interaction of a
supersonic wind from the central source with the interstellar medium is not stationary.
It is schematically shown in Fig.~1. The source continuously injects new particles into
the nebula and the nebula size increases with time. However, in a bounded region of
space near the shock, the flow may be considered to be steady, provided that the nebula
size is much larger than the characteristic size of the shock. This is easy to
understand from simple considerations. Assume that the source injects particles with
characteristic energy~$\epsilon$ into the nebula at a rate $\dot N$. The nebula
size~$R_n$ at constant external pressure~$P_{\mathrm{ext}}$ is then determined by the
relation $\dot N t \epsilon /(4\pi/3 R_n^3)=p_{\mathrm{ext}}$, where $t$ is the source
operation time. We see from this relation that the nebula radius increases as~$t^{1/3}$.
This means that the velocity of the nebula outer rim decreases with time; therefore, the
entire flow may be considered in the limit $t \rightarrow \infty $ as steady with the
boundary condition at infinity $v \rightarrow 0$ for $r \rightarrow \infty$ and
$P=P_{\mathrm{ext}}$. Thus, the condition for applicability of the quasi-stationary
approximation is $R_n \gg r_{\mathrm{sh}}$, where $r_{\mathrm{sh}}$ is the
characteristic radius of the shock front. For the Crab Nebula, $R_n \approx 2$~pc and
the shock radius is $r_{\mathrm{sh}}=0.1$~pc (Kennel and Coroniti~1984). Consequently,
the postshock flow for the Crab Nebula within~$5r_{\mathrm{sh}}$ may actually be
considered to be steady.

\section*{THE INTERACTION OF A HIGHLY ANISOTROPIC WIND WITH THE INTERSTELLAR MEDIUM}

\subsection*{Basic Simplifications}

The problem of the interaction between a supersonic, highly anisotropic plasma flow and
a homogeneous medium has no exact solution so far. Therefore, below, to estimate the
observed effects that must arise during such an interaction, we proceed as follows:
first, we use a highly simplified interaction model to calculate the volume luminosity
of the plerion produced by a Crab-type pulsar; subsequently, we qualitatively consider
how our simplifications affect the results of our calculations by using, in particular,
the results obtained for weak anisotropy in Appendix~A.

We use the following simplifications to estimate the volume luminosity of the Crab
Nebula:

(1) Since the postshock plasma flow is subsonic, with the plasma velocity tending to
zero when moving downstream of the shock, the plasma density along a streamline may be
assumed to be constant (Landau and Lifshitz~1986). The subsonic motion also implies an
approximate equality of the pressure in the plerion. Previously, Begelman~(1992) used
this approximation to describe the postshock flow. Let us estimate the accuracy with
which these conditions may be considered to be satisfied in our specific case. The
pressure variation in the plerion is $\Delta P/P \sim W u^2/P$. Here, $W=4P$ and $u\sim
1/\sqrt{8}$, because the plasma velocity is $v={1\over 3} c$ immediately behind the
shock and then rapidly decreases. Therefore, in the worst case, we have $\Delta P/P \sim
0.5$. This error completely suits us, because below, we are concerned with the
variations in mean particle energy~$\epsilon$ and plasma density~$n$ across a
streamline, which are four orders of magnitude; the pressure is related to these
quantities by $P={1\over 3}n\epsilon$. Against the background of such variations, the
pressure variations of~$50\%$ are of no fundamental importance.

(2) We assume the postshock streamlines to remain radial without bending at the shock
and use the conditions for a perpendicular shock to determine the postshock plasma
parameters.

\subsection*{The Shape of the Shock Front}

To determine the shape of the shock front, we use the shock-adiabat relation~(A.18) for
an oblique shock wave. After the passage of the shock front, the plasma on each
streamline adiabatically decelerates; far from the shock front, its velocity tends to
zero and the plasma pressure comes into equilibrium with the external
pressure~$P_{\mathrm{ext}}$. The relationship between the plasma parameters immediately
after the shock and $P_{\mathrm{ext}}$ is given by the Bernoulli equation for a
relativistic plasma

\begin{equation}
4\gamma_2 {P_2\over n_2}=4{P_{\mathrm{ext}}\over
n_{\infty}}=\gamma_1 mc^2.
\end{equation}
Here, we use the constancy of the Bernoulli integral at the shock front and the
relativistic-plasma approximation~$e=3p$, which holds good behind the shock. Below, the
subscripts~`1' and~`2' denote the preshock and postshock quantities, respectively.

It is easy to find from these relations and from the adiabatic
flow condition~(A.5) that

\begin{equation}
{P_{\mathrm{ext}}\over \gamma_1^2n_1mc^2}={3\over
2}\left({3\over\sqrt{8}}\right)^{{\delta\over
\delta-1}}\gamma_{||}^{{2-\delta\over \delta-1}}.
\end{equation}
The location of the shock front could be determined from this equation if~$\gamma_{||}$
were known. Since the problem is azimuthally symmetric, this quantity is known only at
points on the equatorial plane and on the rotation axis. The shock front crosses them at
a right angle. Below, we consider a simplified case by assuming, for simplicity, that
$\gamma_{||}=1$ everywhere. We then obtain for the shock radius

\begin{equation}
r_{\mathrm{sh}}=\sqrt{{3\over
2}\left({3\over\sqrt{8}}\right)^{{\delta\over \delta-1}}{\gamma_1
mc^2\dot N\over 4\pi P_{\mathrm{ext}} c}}. \label{rshock:Bogovalov_n}
\end{equation}
In the limit $\gamma_m \gg \gamma_0$ of interest, the shock radius
may be assumed to be $r_{\mathrm{sh}}=r_{\mathrm{eq}}|\sin\theta|$
everywhere, except for a narrow interval of angles $\theta <
\sqrt{\gamma_0/\gamma_m}$. This implies that the shock front is
generally a torus whose cross~section is a couple of contacting
circumferences with the centers on the equatorial plane in the
middle of the distance~$r_{\mathrm{eq}}$ from the source to the
shock.

To estimate an  error in using everywhere the
approximation~$\gamma_{||}=1$, we must solve the problem more
accurately. Our analysis, which is beyond the scope of this paper,
shows that the shock front lies at slightly larger distances
than~(10). In addition, a system of two shocks is formed near the
axis for a highly anisotropic wind. Nevertheless, expression~(10)
gives an error in the shock radius within $15\%$. In this paper,
such an accuracy is admissible.

\subsection*{The Formation of a Toroidal Structure}

As follows from the condition of a constant density along the postshock streamline, the
plasma velocity~$v_p$ on each streamline behaves as

\begin{equation}
v_p={1\over 3}c \left({r_{\mathrm{sh}}(\theta)\over r}\right)^2,
\label{vel:Bogovalov_n}
\end{equation}
where $r_{\mathrm{sh}}(\theta)$ is the distance from the pulsar to the shock location at
a given angle~$\theta$. This dependence on~$r$ follows from the conservation of mass
flux for radial plasma motion.

On a given streamline, the dependence of the magnetic field on~$r$ and~$\theta$ follows
from the frozen-in condition. According to this condition, $B/nr=\mathrm{const}$ on a
streamline (Landau and Lifshitz~1982). For the preshock magnetic field, we use the
condition

\begin{equation}
{(B_0{r_0\over r_{\mathrm{sh}}}\sin\theta)^2\over 4\pi
n_1mc^2\gamma_1^2}= \sigma{\gamma_m\sin^2\theta\over \gamma_1}.
\label{b:Bogovalov_n}
\end{equation}
It implies that the ratio of the Poynting flux density to the plasma kinetic energy flux
density is everywhere equal to the same value~$\sigma$, except for a narrow region near
the rotation axis where the toroidal field must vanish. We inserted the additional
factor ${\gamma_m\sin^2\theta\over \gamma}$ in the right-hand part of Eq.~(12) to take
into account this circumstance. It follows from expression~(9) that the preshock
magnetic field is

\begin{equation}
B_1=B_0{r_0\over r_{\mathrm{sh}}}\sin\theta=\sqrt{4\sigma\pi
\gamma_m\gamma_1 n_1mc^2}\sin\theta.
\end{equation}
Given that the magnetic field increases in strength by a factor of~3 after the passage
of a strong shock, we obtain

\begin{equation}
B_{2}=3\sqrt{({4\pi \sigma\gamma_m\over\gamma}{2\over
3}\left({\sqrt{8}\over 3}\right)^{{\delta\over\delta-1}}
P_{\mathrm{ext}})}\sin\theta{r\over r_{\mathrm{sh}}}.
\end{equation}
At angles $\theta \gg \sqrt{\gamma_0/\gamma_m}$, the expression for the magnetic field
takes the form

\begin{equation}
B_{2}=B_{\mathrm{eq}}{r\over r_{\mathrm{sh}}},
\end{equation}
where $B_{\mathrm{eq}}$ is the postshock equatorial magnetic field. We see that the field
immediately behind the shock is everywhere the same, except for a narrow region near the
rotation axis. Outside this region, the field more rapidly increases with distance from
the pulsar at high latitudes. The field linearly increases until the magnetic energy
density becomes equal to the plasma energy density. Subsequently, the field begins to
decrease with increasing~$r$ (Kennel and Coroniti~1984). The fact that the linear
dependence extends to~$5 r_{\mathrm{sh}}$ (Kennel and Coroniti~1984), within which the
toroidal structure is formed, will suffice.

To calculate the synchrotron radiation, we assume, as in Kennel and Coroniti~(1984),
that the following power-law particle spectrum is formed behind the shock:

\begin{equation}\label{f1:Bogovalov_n}
F(\gamma,r=r_{\mathrm{sh}})=
A\left({\gamma\over\gamma_{\min}}\right)^{-\alpha}
\eta(\gamma-\gamma_{\min})\eta(\gamma_{\max}-\gamma),
\end{equation}
where $A$ is the normalization factor determined from the particle injection rate on a
given streamline, $\gamma_{min}$ is the minimum particle energy in the spectrum,
$\gamma_{max}$ is the maximum particle energy in the spectrum, and $\eta(x)$ is the step
function equal to unity at $x \ge 1$ and zero at $x < 1$.

The mean energy of the chaotic particle motion in this spectrum must correspond to the
mean postshock particle energy determined from the shock adiabat. It follows from this
condition that

\begin{equation}
\gamma_{\min}={\gamma_1(\theta)\over
\sqrt{2}}{(\alpha-2)\over(\alpha-1)}
\end{equation}
Here, we use the fact that although the cutoff energy of the spectrum, $\gamma_{\max}
\approx 5\times 10^9$ (Atoyan and Aharonian~1996), is finite, it is much larger than
$\gamma_{\min} \le 5\times 10^5$.

The plasma moves behind the shock as a whole at velocity~(11). The evolution of the
particle spectrum during this motion is described by the transport equation

\begin{equation}
v_p{\partial F\over \partial r}=-{\partial\over \partial\gamma }
{\dot{\gamma}} F \label{el:Bogovalov_n}
\end{equation}
The rate of change in the Lorentz factor of the chaotic particle motion, $\dot\gamma$,
is generallyly determined by synchrotron losses and plasma heating thrgh adiabatic
compression during the wind deceleration. In our case, the adiabatic changes in particle
energy can reach~$24\%$ of the particle energy. However, we disregard these changes
here, because this is a clear excess of the accuracy under our assumptions about the
postchosk plasma dynamics.

The solution of Eq.~(18), provided that the distribution function at the shock matches
function~(16), is

\begin{eqnarray}\label{spectr:Bogovalov_n}
f(\gamma,r,\theta)={(1-\gamma G)^{(\alpha-2)}(\alpha-1)\over
\gamma_{\min}} \left({\gamma\over
\gamma_{\min}}\right)^{-\alpha}\times\\
\nonumber\times\eta\left({\gamma\over 1-\gamma
G}-\gamma_{\min}\right)
\eta\left(\gamma_{\max}-{\gamma\over(1-\gamma
G)}\right)n(r,\theta).
\end{eqnarray}
In this expression, the function

\begin{equation}
G(r,\theta)={2\over 5} r_{\mathrm{sh}}(\theta)\left({e^2\over
mc^2}\right)^2 {B_{\mathrm{eq}}^2\over mc^2} \left(\left({r\over
r_{\mathrm{sh}}(\theta)}\right)^5-1\right)
\end{equation}
describes the degradation of the particle energy through synchrotron losses. The
function~$ n(r,\theta)$ is the emitting-electron density in the observer's frame of
reference. The electron density in the plerion on the streamline with~$\theta$ follows
from expression~(6):

\begin{equation} \label{n:Bogovalov_n}
n(r,\theta)= {3 \dot N\over 4\pi c r_{\mathrm{sh}}^2(\theta)}.
\end{equation}
Here, we took into account the fact that the density at the shock increases by a factor
of~3. The synchrotron flux is determined by the convolution of spectrum~(19) with the
spectral distribution of the radiation from an individual electron (Landau and
Lifshitz~1973). The input parameters for our calculations are the particle injection
rate $\dot N=5\times 10^{38}$~part.~s$^{-1}$, the maximum Lorentz factor of the wind at
the equator $\gamma_m=4.5\times 10^6$, and the shock location at the equator; we
correlated the latter with the inner ring of the toroidal structure, which is located at
$r_{\mathrm{eq}}=4.3\times 10^{17}$~cm corresponding to a distance of~$14''$ (Weisskopf
\emph{et al.}~2000). Actually, this implies that the external pressure~$P_{\mathrm{ext}}$
was chosen so that the shock at the equator was located on the inner ring of the toroidal
structure.

The results of our calculations are presented in Fig.~2. This figure shows the plerion
volume luminosity at photon energy 400~eV, the characteristic energy at which the Chandra
observations are carried out (Weisskopf \emph{et al.}~2000). No integration along the
line of sight was performed. All geometric sizes were reduced to the pulsar distance,
2~kpc. Here, only the plerion cross~section in the poloidal plane is shown. We see that
the spatial distribution of the volume luminosity has the shape of a torus with the
characteristic sizes corresponding to the observed ones. The radiation reaches the
highest intensity at a distance of~$\approx40''$, where the second outer ring in the
Crab Nebula is located. The intensity of the synchrotron radiation behind the shock is
known to gradually rise (Kennel and Coroniti~1984). This rise results from a linear
increase in the magnetic field behind the shock, which takes place only if the preshock
wind was weakly magnetized. The further rapid decline in the radiation at distances
$>60''$ is attributable to fast synchrotron electron cooling in the growing magnetic
field. We see that although our calculations are definitely incorrect for $ r >
5R_{\mathrm{sh}}$, this is of no importance. When the electrons reach this region, they
have already cooled down to an extent that they do not produce detectable radiation in
the Chandra spectral range.

In our calculations, we failed to obtain the bright inner toroidal ring. It may well be
that this brightening cannot be explained in terms of magnetic hydrodynamics in
principle and that it is attributable to the structure of a collisionless shock, which
must be investigates separately (see~Gallant~1992).

Across the equatorial plane, the agreement with the observed distribution is slightly
worse. The distribution of the calculated brightness across the equator is broader than
that of the observed one. It may well be that distributions (5)--(7) do not faithfully
describe the preshock wind parameters. The energy and plasma fluxes in the wind may be
more concentrated toward the equator than follows from the theory. In our view, however,
it is too early to draw this conclusion. The observed disagreement may be entirely
attributable to our approximations. We assumed the shock to be perpendicular to the
streamline everywhere. This is not true outside the equatorial plane, where the shock
inclination to the streamline decreases while $\gamma_{||}$ increases. As the angle of
incidence of the shock decreases, the mean energy of the chaotic particle motion after
the shock passage also decreases, as is clearly seen from the relation for the shock
adiabat of an oblique relativistic shock. Therefore, the volume intensity must decline
with height above the equator faster than in our case.

We disregarded the fact that the streamlines must bend toward the equator at the shock,
as is the case for weak anisotropy under consideration. Allowing for this circumstance
will cause an increase in the brightness near the equator and its faster decrease under
it. Finally, we assumed the streamlines after the shock passage to remain straight. In
fact, as our calculations for low anisotropy show, the streamlines continue to approach
the equator, at least in some preshock region, even after the shock passage. Thus, all
our most important approximations used to estimate the volume luminosity of the plerion
produced by the Crab pulsar lead to the same effect: a broader distribution of the
radiation intensity than must be in the case of a more accurate calculation. Therefore,
so far we have no reason to believe that distributions (5)--(7) inaccurately describe the
preshock parameters of the wind from the Crab pulsar.

\subsection*{Jet-like Features}

In the model that is a simple generalization of the model by Kennel and Coroniti~(1984),
we failed to obtain something similar to the bright jet-like features observed in the
Crab Nebula. However, noteworthy is one circumstance that has a direct bearing on the
observed jets. As we already pointed out above, the plasma pressure in the plerion is
roughly the same, because the flow is subsonic. The relativistic-plasma pressure is given
by the relation $p={1\over 3} n(r,\theta )\epsilon$, where $\epsilon$ is the mean
particle energy. Since it is proportional to the Lorentz factor of the preshock plasma
$\gamma_1$, the plasma density in the plerion is

\begin{equation}
n(r,\theta)= n_{\mathrm{eq}}{\gamma_0+\gamma_m\over
\gamma_0+\gamma_m \sin^2\theta}, \label{dens:Bogovalov_n}
\end{equation}
where $n_{\mathrm{eq}}$ is the equatorial plasma density. Density~(22) appears in
expression~(19) for the electron spectrum. We see from this expression that
$n_{\mathrm{axis}}/n_{\mathrm{eq}}=\gamma_m/\gamma_0$. In standard models, $\gamma_0
\approx 200$ and $\gamma_m \approx 3\times10^6$, implying that the plasma density in the
plerion near the rotation axis is approximately a factor of 15~000 higher than the
equatorial plasma density. It is important to note that this fact is a direct result of
the energy distribution in the wind~(5) and depends weakly on any other factors. Thus, if
the luminosity were simply proportional to the plasma density, then we would observe
only an extremely bright jet from the pulsar in the plerion with a barely visible torus
against its background. However, this is not the case. The increase in density toward the
rotation axis causes no brightening, because this increase is accompanied by the
simultaneous decrease in particle energy. As a result, the volume intensity decreases
toward the rotation axis.

Thus, the formation of dense but relatively cold jet-like features in the plerion along
the rotation axis necessarily follows from the theory of the interaction between the
Crab pulsar anisotropic wind and the interstellar medium. There is only one problem:
through which processes these features can become bright enough to be observable. A
possible solution could be the assumption that some particle acceleration takes place
not only at the shock but also in the entire volume of the plerion. This assumption
seems reasonable (Begelman~1998) and is supported by the detection of gamma-rays with
energy above 50~TeV from the Crab Nebula (Tanimori \emph{et al.}~1998). If we assume
that an additional weak particle acceleration takes place in the entire volume of the
nebula, the spectral shape of the accelerated particles is the same everywhere, and the
number of accelerated particles is proportional to the plasma density in the plerion,
then a second radiation component proportional to the particle density in the nebula
emerges. We see from Fig.~2 that in this case, the second radiation component manifests
itself in the form of bright jets, with the fraction of the accelerated particles being
$2\times 10^{-8}$ of their local density.

\section*{CONCLUSIONS}

Based on the standard theory by Kennel and Coroniti~(1984), we
have shown that the principal features of the observed X-ray
structure in the Crab Nebula can be naturally explained by taking
into account anisotropy of the energy flux in the winds from radio
pulsars. For our calculations, we used the fact that the energy
flux density in the pulsar wind is proportional to $\sin^2\theta$.
This dependence follows from the expression for the Poynting
flux~$P_y$ for axisymmetric winds, $P_y\sim r^2\sin^2\theta
B_p^2$, provided that the poloidal magnetic field~$B_p$ pulled by
the wind from the source is isotropic. Such a dependence remains
valid even for an obliquely rotating source with a uniform
magnetic field, although, in general, the flow is not steady and
axisymmetric in this case (Bogovalov~1999). For a nonuniform
(in~$\theta$) poloidal magnetic field, the $\theta$ dependence of
the energy flux density can change. However, it is easy to
understand that the energy flux density is at a minimum on the
rotation axis and reaches a maximum at the equator, irrespective
of the rotation angle and at any reasonable distribution of the
poloidal magnetic field in~$\theta$. This is because the Poynting
flux density is proportional to the toroidal magnetic
field~$B_{\varphi}$. However, $B_{\varphi}=0$ always on the axis.
Therefore, $P_y=0$ on the rotation axis and $P_y$ can only
increase when moving away from the axis.

The breakdown of azimuthal uniformity of the poloidal magnetic field leads to the
additional generation of magnetosonic waves (but not magnetodipole radiation) in the
wind. However, the energy flux in them is also proportional to the combination
$r^2\sin^2\theta\langle B_p^2\rangle$ (at least for a small wave amplitude;
Bogovalov~2001a). Therefore, irrespective of the rotation angle, $Py$ always reaches a
maximum at the equator, except for the exotic case where~$B_p$ decreases toward the
equator faster than $\sin^{-1}\theta$. In other words, the concentration of the energy
flux density toward the equator is apparently a common property of the pulsar winds.

The formation of a bright X-ray torus is a direct result of this common property of the
pulsar winds. The interaction between a wind with such anisotropy and the interstellar
medium also inevitably gives rise to cold subsonic (jet-like) flows in the plerion along
the rotation axis whose density is almost four orders of magnitude higher than the
equatorial density. In the standard theory, these flows are invisible in X-rays, because
the particle energy is too low to produce detectable synchrotron radiation. However, if
we assume the additional acceleration of a mere $10^{-8}$ fraction of the particles at
each point of the plerion, then the radiation from the jet-like features becomes
comparable in intensity to the the torus radiation and the overall morphology of the
plerion becomes similar to that observed on the Chandra observatory. It would be natural
to assume that similar features detected around the Vela pulsar (Pavlov \emph{et
al.}~2001) and in the supernova remnant G$0.9+01$ (Gaensler~2001) can be explained in a
similar way, because the anisotropy in the pulsar wind of the type discussed here must
be formed in all pulsars.

\section*{ACKNOWLEDGMENTS}

This study was supported in part by the joint INTAS--ESA grant no.~99-120 and as part of
the project ``Universities of Russia--Basic Science'' (registration no.~015.02.01.007).


\appendix
\def\appendixname{APPENDIX}%
\section{A WEAKLY ANISOTROPIC WIND}

The actual pulsar wind has a strong latitudinal dependence of the
particle energy. The ratio~$\gamma_m/\gamma_0$ is of the order
of~$10^4$. However, for a qualitative understanding of the
interaction between an anisotropic wind and the ambient medium, it
is of interest to consider this interaction for a weak anisotropy.
This problem is valuable in that its solution can be obtained
analytically.

As was already pointed out above, the approximation of a
hydrodynamic interaction is invoked to describe the plasma flow.
In addition, the problem is axisymmetric. Let us introduce the
stream function~$\psi$. It is related to the physical quantities
by

\begin{equation}
nu_{r} =
\frac{1}{r\sin\theta}\frac{\partial\psi}{r\partial\theta},
\end{equation}
\begin{equation} \label{utheta:Bogovalov_n}
nu_{\theta}=-\frac{1}{r \sin\theta}\frac{\partial\psi}{\partial
r},
\end{equation}
where $n$ is the intrinsic particle number density and
$u_{r},u_{\theta}$ are the corresponding components of the
four-velocity. It is convenient to use the equations for the
stream function in spherical coordinates (Beskin~1997)

\begin{eqnarray}\label{main:Bogovalov_n}
\hspace{-5pt}(u_{s}^{2}-u_{r}^{2})\psi_{rr}+(u_{s}^{2}-u_{\theta}^{2})\frac{\psi_{\theta
\theta}}{r^{2}}-2 u_{r}u_{\theta}\frac{\psi_{r\theta}}{r}-
u_{s}^{2} n u_{r}\cos\theta +nu_{\theta}\sin\theta(2u_{r}^{2}+
u_{\theta }^{2}) \hspace{-1pt}=\\ \nonumber
 \hspace{10pt}=(n r \sin
\theta)^{2} \hspace{-1pt}\left[u_{s}^{2} \gamma^{2} \frac{d\ln
A}{d \psi} + \frac{\delta - 1}{\delta} (u_{s}^{2}-u_{r}^{2})
\frac{d\ln S}{d\psi}\right],
\end{eqnarray}
where $u_{s}^{2}=\frac{\delta-1}{2-\delta}$ and $\delta=4/3$ is
the polytropic index of the relativistic plasma,

\begin{equation} \label{S:Bogovalov_n}
S(\psi)=\frac{p}{n^{\delta}}.
\end{equation}
Since the motion is adiabatic, $S$ is conserved along a streamline
and depends only on~$\psi$. The ratio of the Bernoulli integral
and~$S$ is

\begin{equation} \label{A:Bogovalov_n}
A=\frac{\delta}{\delta-1}\frac{\gamma w}{n S(\psi)}
\end{equation}
It is also conserved along a streamline. Here, $w$ is the thermal
function per~particle in the intrinsic coordinate system and $p$
is the intrinsic pressure. In the ultrarelativistic case, $w$ and
$p$ are related by the equation of state
$w=\frac{\delta}{\delta-1}p$. For weak anisotropy, we solve
Eq.~(A.3) using the perturbation theory by assuming that the
Lorentz factor of the wind depends on the polar angle as

\begin{equation}
\gamma=\gamma_{w}(1+\alpha\sin^{2}\theta)
\label{gamma0:Bogovalov_n},
\end{equation}
where $\alpha$ is a small dimensionless parameter and $\gamma_w$
is the Lorentz factor of the wind.

In the zero approximation, the flow is spherically symmetric,
$u_{\theta}=0$, and the following identities hold: $\frac{d
A}{d\psi}=0$ and $\frac{d S}{d\psi}=0$. The right-hand side of
Eq.~(A.3) becomes zero and it takes the form

\begin{equation} \label{main0:Bogovalov_n}
u_{s}^{2}\psi_{0\theta \theta}- u_{s}^{2} n u_{r} \cos \theta =0.
\end{equation}
The solution of Eq.~(A.7) is

\begin{equation} \label{null:Bogovalov_n}
\psi_{0}=n_{\mathrm{sh}}u_{\mathrm{sh}}r_{\mathrm{sh},0}^{2}(1-\cos\theta),
\end{equation}
where $r_{\mathrm{sh},0}$ is the radius of the spherical shock
wave and $n_{\mathrm{sh}},u_{\mathrm{sh}}$ are the density and the
radial component of the four-vector immediately behind the shock.
According to Landau and Lifshitz~(1986),
$u_{\mathrm{sh}}=\frac{c}{\sqrt{8}}$. To solve the equation in the
first order of the perturbation theory, we represent the stream
function as

\begin{equation} \label{psi:Bogovalov_n}
\psi=n_{\mathrm{sh}}u_{\mathrm{sh}}r_{\mathrm{sh},0}^{2}\left(1-\cos\theta+\alpha
f(r,\theta)\right),
\end{equation}
where $f(r,\theta)$ is the correction to the first approximation.
The equation for it is

\begin{eqnarray}\label{pert:Bogovalov_n}
(u_{s}^{2}-u_{0}^{2})f_{rr}+\frac{1-\eta^{2}}{r^{2}}u_{s}^{2}f_{\eta\eta}
 -2 \frac{u_{0}^{2}}{r} f_{r }=\\ \nonumber
 = -\frac{(n_{0}
r)^{2}}{\left(n_{\mathrm{sh}}r_{\mathrm{sh}}^{2}u_{\mathrm{sh}}\right){2}}(1-\eta^{2})
 \left[u_{s}^{2} \gamma^{2} \frac{d\ln A(\psi)}{d \eta} +
\frac{\delta - 1}{\delta} (u_{s}^{2}-u_{0}^{2}) \frac{d\ln
S(\psi)}{d\eta}\right],
\end{eqnarray}
where $\eta=cos\theta$ and $u_{0}$, $n_{0}$, $\gamma$ are the
solution behind the shock in the zero approximation.

To determine $A$ and $S$, we use the shock-adiabat relations for
an oblique relativistic shock wave. These are derived from Landau
and Lifshitz~(1986) using the Lorentz transformations

\begin{equation} \label{shock1:Bogovalov_n}
v_{1,\bot}=\sqrt{1-v_{\|}^{2}}\left[\frac{(p_{2}-p_{1})(e_{2}+p_{1})}{(e_{2}-e_{1})(e_{1}+p_{2})}\right]^{1/2},
\end{equation}
\begin{equation} \label{shock2:Bogovalov_n}
v_{2,\bot}=\sqrt{1-v_{\|}^{2}}\left[\frac{(p_{2}-p_{1})(e_{1}+p_{2})}{(e_{2}-e_{1})(e_{2}+p_{1})}\right]^{1/2},
\end{equation}
where $p$ is the pressure and $e$ is the internal energy. The
quantities with the subscripts~$1$ and $2$ describe the preshock
and postshock states of the wind, respectively. $v_{\|}$ and
$v_{\bot}$ are the tangential and normal velocity components,
respectively.

Since the thermodynamic state of the matter in the preshock region
corresponds to a zero temperature,

\begin{equation}
p_{1}=0,e_{1}=mc^{2}n_{1}.
\end{equation}
According to~(21, A.6), the density can be represented as

\begin{equation} \label{n1:Bogovalov_n}
n_{1}=(1-\alpha\sin^{2}\theta)\frac{n_{w}}{r^2},\\
 n_w={\dot N \over 4\pi v_p \gamma_w}.\nonumber
\end{equation}

The function~$r_{\mathrm{sh}}(\theta)$ can be represented as a
series of Legendre polynomials~$P_m$,

\begin{equation}
r_{\mathrm{sh}}(\theta)=r_{\mathrm{sh},0}\left(1+\alpha
\sum_{m}R_{m}P_{m}(\cos\theta)\right),
\end{equation}
In front of the shock, we then have

\begin{equation}
n_{1}=\frac{n_{w}}{r_{\mathrm{sh},0}^{2}}
\left(1-\alpha\sin^{2}\theta-2\alpha
\sum_{m}R_{m}P_{m}(\cos\theta)\right),
\end{equation}
\begin{equation}
e_{1}=\frac{mc^{2}n_{w}}{r_{\mathrm{sh},0}^{2}}
\left(1-\alpha\sin^{2}\theta-2\alpha
\sum_{m}R_{m}P_{m}(\cos\theta)\right).
\end{equation}

In the ultrarelativistic case, the following relations hold:
$e_{2}\gg e_{1}$ and $p_{2}\gg e_{1}$. They allow (A.11, A.12) to
be simplified. For~$\delta=4/3$, we have

\begin{equation}
p_{2}=\frac{2}{3}\frac{\gamma_{1}^{2}}{\gamma_{\|}^{2}}e_{1},\quad
\gamma_{2}^{2}=\frac{9}{8}\gamma_{\|}^{2}, \label{a18:Bogovalov_n}
\end{equation}
where $\gamma_{\|}=\frac{1}{\sqrt{1-v_{\|}^{2}}}$.

The postshock plasma density can be calculated by using the
Bernoulli integral

\begin{equation}
n_{2}=n_{1}\frac{\gamma_{2}}{\gamma_{1}}\frac{\delta}{\delta-1}\frac{p_{2}}{e_{1}}.
\end{equation}

For weak anisotropy, $v_{\|}\ll 1$. We see from the geometry shown
in Fig.~3 that

\begin{equation} \label{gamma_par:Bogovalov_n}
\frac{1}{\gamma_{\|}} \approx 1-{1\over 2}v_1^2\sin^2\beta \approx
1-\frac{v_1^{2}}{2}\left(\frac{r_{\mathrm{sh}}'(\theta)}{r_{\mathrm{sh}}(\theta)}\right)^{2}.
\end{equation}
Since $r^{'}_{\mathrm{sh}} \sim \alpha$, the first corrections to
$\gamma_{||}$ begin with the terms proportional to~$\alpha^2$. In
the first order of smallness in~$\alpha$, we may assume that
$\gamma_{||}=1$. The change in the postshock thermodynamic
quantities when the flow isotropy breaks down results only from a
change in the spatial location of the shock front. To a first
approximation, the change in the inclination of the shock surface
is negligible.

In that case,

\begin{equation} \label{p2:Bogovalov_n}
p_{2}=\frac{2}{3}\frac{n_{w}}{r_{\mathrm{sh},0}^{2}}\gamma_{w}^{2}mc^{2}
\left(1+\alpha\sin^{2}\theta-2\alpha
\sum_{m}R_{m}P_{m}(\cos\theta)\right),
\end{equation}
\begin{equation}
\gamma_{2}^{2}=\frac{9}{8},
\end{equation}
\begin{equation}
n_{2}=3\gamma_{0}n_{w}\left(1-2\alpha
\sum_{m}R_{m}P_{m}(\cos\theta)\right).
\end{equation}
Hence, we have for $A$ and $S$

\begin{equation}
S=S_{0}\Bigg(1+\alpha\sin^{2}\theta {+}{}\\
\nonumber{}+\left.2/3\alpha \sum_{m}R_{m}P_{m}(\cos\theta)\right),
\end{equation}
\begin{equation}
A=A_{0}\left(1-2/3\alpha \sum_{m}R_{m}P_{m}(\cos\theta)\right),
\end{equation}
where $A_{0}$ and $S_{0}$ are the constants that are of no
interest.

Substituting the expressions for $A$ and $S$ in relation~(A.10)
yields

\begin{eqnarray}\label{pertn:Bogovalov_n}
(u_{s}^{2}-u_{0}^{2})f_{rr}+\frac{1-\eta^{2}}{r^{2}}f_{\eta\eta}
-2 \frac{u_{0}^{2}}{r} f_{r } =\\ \nonumber =\frac{(n_{0}
r)^{2}}{\left(n_{\mathrm{sh}}r_{\mathrm{sh}}^{2}u_{\mathrm{sh}}\right){2}}(1-\eta^{2})
\left[2\eta\frac{u_{s}^{2}-u_{0}^{2}}{4}\right.
-\left.2/3\left(u_{s}^{2} \gamma^{2}-\frac{u_{s}^{2}-u_{0}^{2}}{4}
\right) \sum_{m}R_{m}P_{m}'(\eta)\right].
\end{eqnarray}

We seek a solution to Eq.~(A.26) in the form

\begin{equation} \label{f:Bogovalov_n}
f(r,\eta)=\sum_n Q_n(\eta)f^{(m)}(r),
\end{equation}
where $Q_m(\eta)$ are the eigenfunctions of the operator

\begin{equation}
(1-\eta^{2})Q_m''(\eta)=-m(m+1)Q_m(\eta);
\end{equation}
These are

\begin{equation}
Q_m(\eta)=(1-\eta^{2})P_m'(\eta).
\end{equation}

Substituting~(A.27) in (A.26) yields

\begin{eqnarray}\label{pertx:Bogovalov_n}
(u_{s}^{2}-u_{0}^{2})f^{(m)}_{xx}-\frac{m(m+1)}{x^{2}}u_{s}^{2}f^{(m)}
-2 \frac{u_{0}^{2}}{x} f^{(m)}_{x}=\\ \nonumber ={1\over
u_{0}^{2}x^{2}}\left[\frac{2}{3}\delta_{2,m}\frac{u_{s}^{2}+u_{0}^{2}}{4}
+2/3\left(u_{s}^{2} \gamma^{2}-\frac{u_{s}^{2}-u_{0}^{2}}{4}
\right) R_{m}\right],
\end{eqnarray}
where $x=r/r_{\mathrm{sh},0}$.

The function~${u_0}$ was numerically determined from the equations
for a polytropic, spherically symmetric postshock flow. This
function depends only on one variable~$x$. Since the first
boundary condition at the shock front is the continuity of~$\psi$,

\begin{equation} \label{bcond0:Bogovalov_n}
f^{(m)}(1)=0.
\end{equation}
The second boundary condition follows from the shock-adiabat
relations for an oblique shock wave. Behind the shock,

\begin{equation}
n_{2}u_{2
\theta}=-\alpha\frac{r_{\mathrm{sh}}^{2}n_{\mathrm{sh},0}u_{\mathrm{sh}}}{r
\sin\theta}\frac{\partial f}{\partial r}.
\end{equation}
We see from Fig.~3 that at $\varphi \ll 1$ and $\beta \ll 1$,

\begin{equation} \label{phi:Bogovalov_n}
\frac{u_{2\theta}}{u_{2r}}=\varphi=-\frac{\alpha}{\sin\theta
x}\frac{\partial f}{\partial x},\quad
\varphi+\beta=\frac{v_{\|}}{v_{2\bot}}.
\end{equation}
In the first order of smallness in~$\alpha$, we have

\begin{equation} \label{all:Bogovalov_n}
v_{2\bot}=\frac{1}{3},\quad
v_{\|}=\beta=\frac{r_{\mathrm{sh}}'(\theta)}{r_{\mathrm{sh}}(\theta)}.
\end{equation}
It follows from~(A.33, A.34) at $x=1$ that

\begin{equation}
\frac{\partial f}{\partial x}=2(1-\eta^{2})\sum_{m}R_m
\frac{dP_m(\eta)}{d\eta},
\end{equation}
and the second boundary condition for $f^{(m)}$ takes the form

\begin{equation} \label{bcond1:Bogovalov_n}
\frac{\partial f^{(m)}}{\partial x}\mid_{x=1}=2R_{m}.
\end{equation}

The general solution of Eq.~(A.30) with the boundary
conditions~(A.31, A.36) is

\begin{equation}
f^{(m)}=\delta_{m,2}f^{(m)}_{1}+ R_{m}f^{(m)}_{2},
\label{family:Bogovalov_n}
\end{equation}
where $f^{(m)}_{1}$ is the solution of the inhomogeneous
equation~(€.30) with the right-hand side ${1\over
u_{0}^2x^2}\frac{2}{3}\frac{u_{s}^{2}-u_{0}^{2}}{4}$ and the
boundary conditions $ f^{(m)}_{i}(1)=0, \quad \frac{\partial
f^{(m)}_{i}}{\partial x}\mid_{x=1}=0$. $f^{(m)}_{2}$ is the
solution of the same equation with the right-hand side ${1\over
u_{0}^2x^2}{2\over 3}\left(u_{s}^{2}
\gamma^{2}-\frac{u_{s}^{2}-u_{0}^{2}}{4} \right)$ and the boundary
conditions $f^{(m)}_{2}(1)=0, {\partial f^{(m)}_{2}\over \partial
x}=2$. Since the functions $f^{(m)}_{1}$ and $f^{(m)}_{2}$
increase with distance~$x$ faster than~$x^4$, $R_m\rightarrow 0$
when $x_{\infty} \rightarrow \infty$ under any boundary condition.
Therefore, for all $m \neq 0$ and for $m \neq 2$, we may assume
that $R_m =0$. This condition physically means that all obstacles
that can be in the flow lie far enough from the shock for their
influence on the flow near the shock to be ignored.

Figure~4 shows a family of solutions~(A.37) for various~$R_2$. All
these solutions satisfy the boundary conditions at the shock. To
single out the only solution, we must formulate the boundary
conditions far from the shock. It is easy to verify that if there
are obstacles in the postshock flow, then the postshock
perturbation grows faster than~$x^4$. In this case, the condition
that there are no additional obstacles behind the shock in the
presence of anisotropy is the requirement that the perturbation
should grow with~$x$ more slowly than~$x^4$, for example,
as~$x^3$. This will take place if the right-hand side of
Eq.~(A.30) tends to zero as $x$ increases. In turn, it is easy to
find that the right-hand side of this equation tends to zero for
$R_2=-{1\over 3}$. The solution for this value of~$R_2$ is
indicated in Fig.~4 by the solid line. It is the separatrix that
separates the solutions tending to $+\infty$ from the solutions
tending to $-\infty$ for $x\rightarrow \infty$.

Although the expansion of~$f$ includes the terms with $m > 1$, we
must also determine~$R_0$, because this term appears in the
expansion of $r_{\mathrm{sh}}$ in terms of Legendre polynomials.
This coefficient can be easily determined from the condition for
the shock location being constant on the rotation axis. The shock
front then takes the form

\begin{equation}
r_{\mathrm{sh}}=r_{\mathrm{sh},0}\left(1+\alpha\left({1\over
3}P_0-{1\over 3}P_2\right)\right)
=r_{\mathrm{sh},0}\left(1+{\alpha\over 2}\sin^2\theta\right).
\end{equation}
In this solution, the postshock pressure~$p_2$ is constant along
the front and the density varies as

\begin{equation}
n_2=3\gamma_w n_w(1-\alpha\sin^2\theta).
\end{equation}
Accordingly, the mean energy of the chaotic particle motion behind
the shock front $\epsilon_2=3p_2/n_2$ is

\begin{equation}
\epsilon_2={2\over 3}\gamma_w mc^2(1+\alpha\sin^2\theta).
\end{equation}
Thus, as the energy flux density increases along the equator, the
shock front is extended along the equatorial plane; the pressure
along the front is constant, to a first approximation. The
postshock particle density at the equator is lower and their mean
thermal energy is higher than those for the particles along the
rotation axis. Figure 5 shows streamlines in the flow for
$\alpha=0.25$. We see that, in addition to this, the increase in
the energy flux density along the equator also causes the bending
of the streamlines toward the equator. Initially, this takes place
at the shock, where, according to the conditions for an oblique
shock wave, the streamlines are pressed to the shock.
Subsequently, the streamlines continue to be pressed to the
equator. Of course, we cannot talk about the behavior of the
solutions far from the shock front, where our solution can yield a
qualitatively incorrect result. However, we see that there is one
tendency at small distances from the shock front: the bending of
the streamlines toward the equator behind the shock.


Translated by V. Astakhov

\pagebreak
\begin{figure}
\plotone{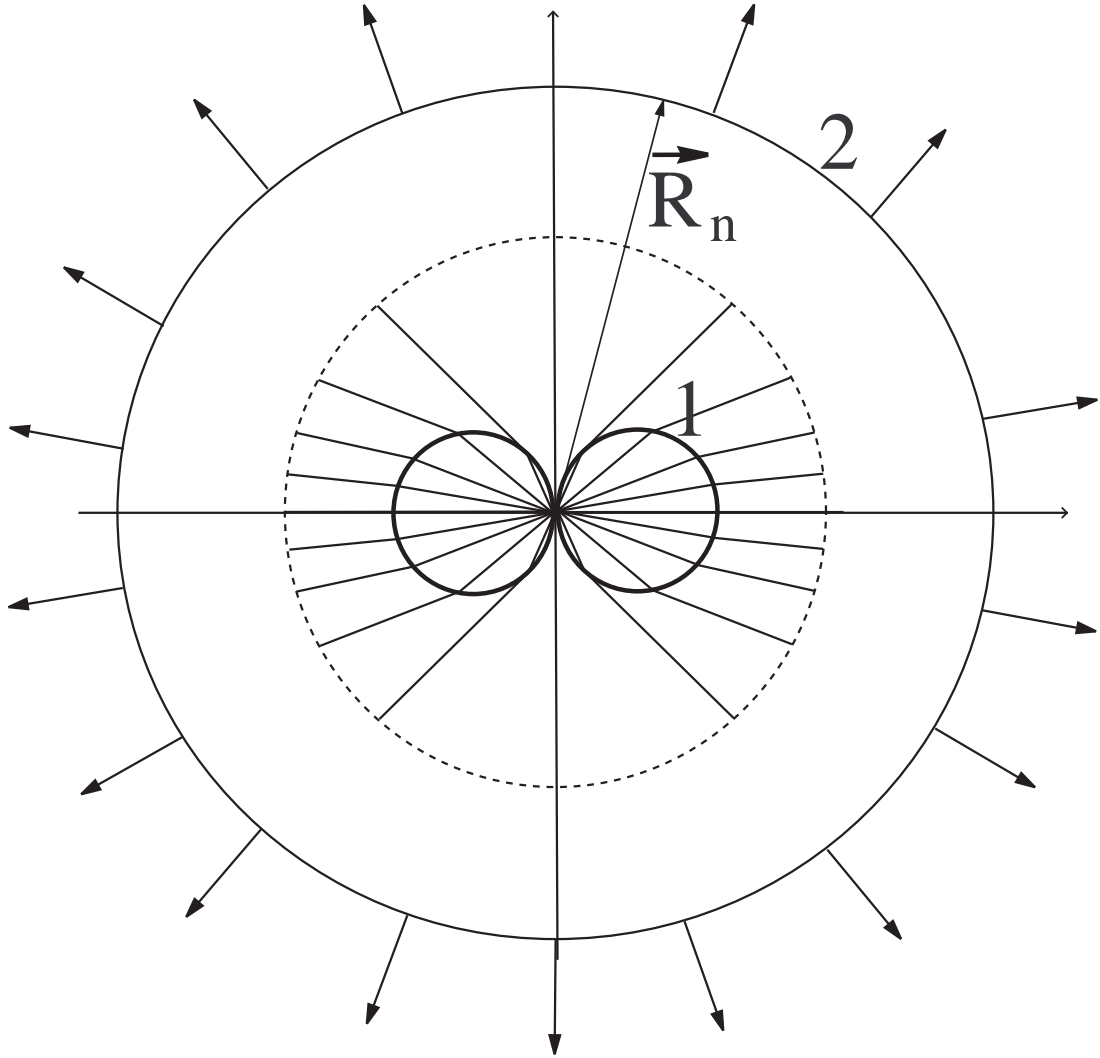}

\caption{A schematic image of the interaction between the pulsar
wind and the interstellar medium. Because of the increase in the
energy flux density in the wind toward the equator, the shock
(\emph{1}) takes the form of a torus. The streamlines bend at the
shock front and are pressed to the equator. The region between the
shock and the contact discontinuity (\emph{2}) is filled with
relativistic particles emitting synchrotron radiation. The dotted
line marks the $r\ll R_{n}$ region, where the plasma flow may be
considered to be quasi-steady \hfill}
\end{figure}
\clearpage
\begin{figure}
\centerline{\psfig{file=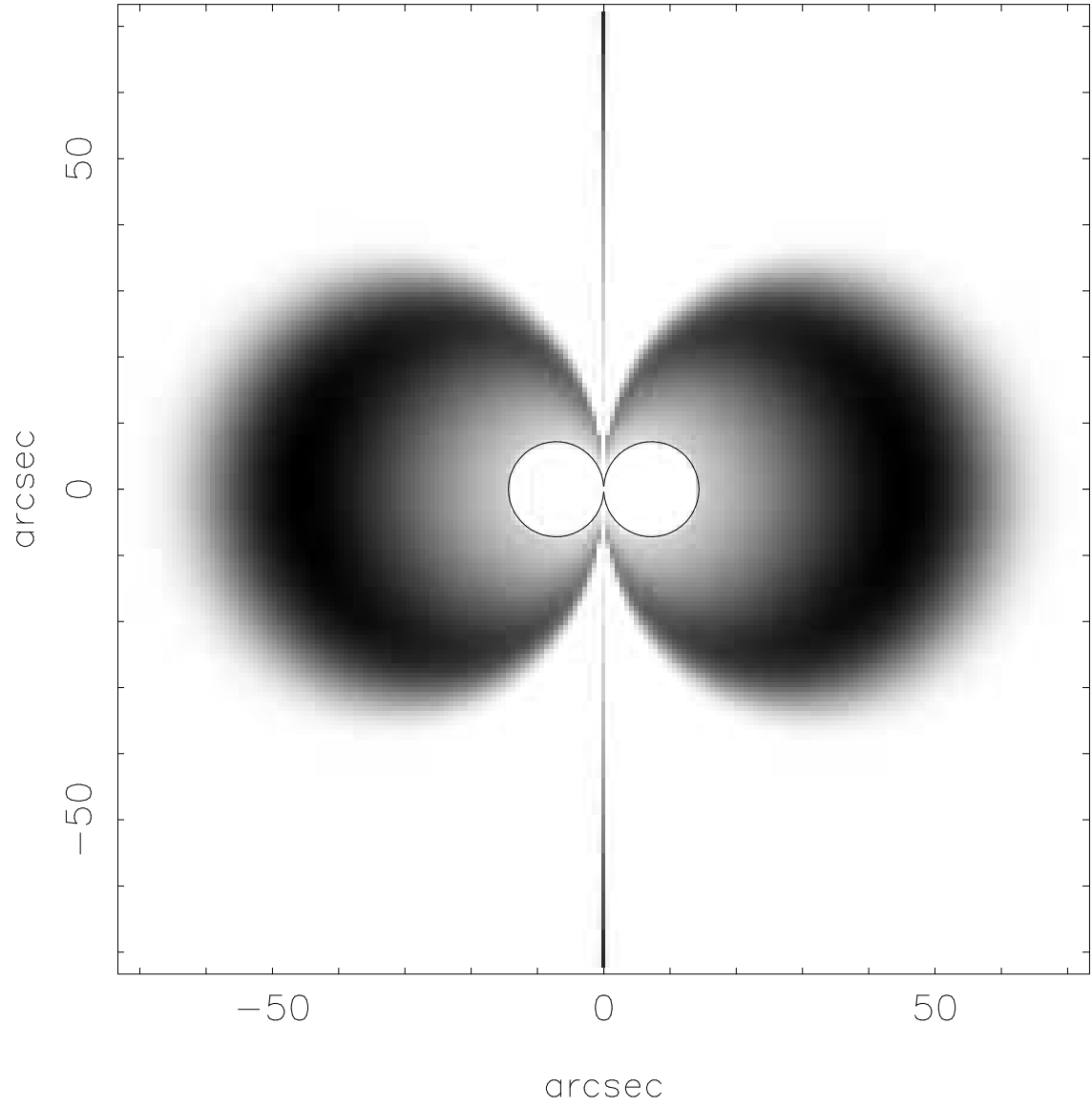,width=150mm,angle=270}}
\caption{The distribution of the volume emissivity of synchrotron
radiation in the plerion produced by the wind from the
pulsar~PSR~$0531+21$. The cross~section of the shock is similar in
shape to two contacting circumferences. The shock location at the
equator is taken to be~$14''$, in accordance with the observed
location of the torus inner ring. \hfill}
\end{figure}
\begin{figure}
\plotone{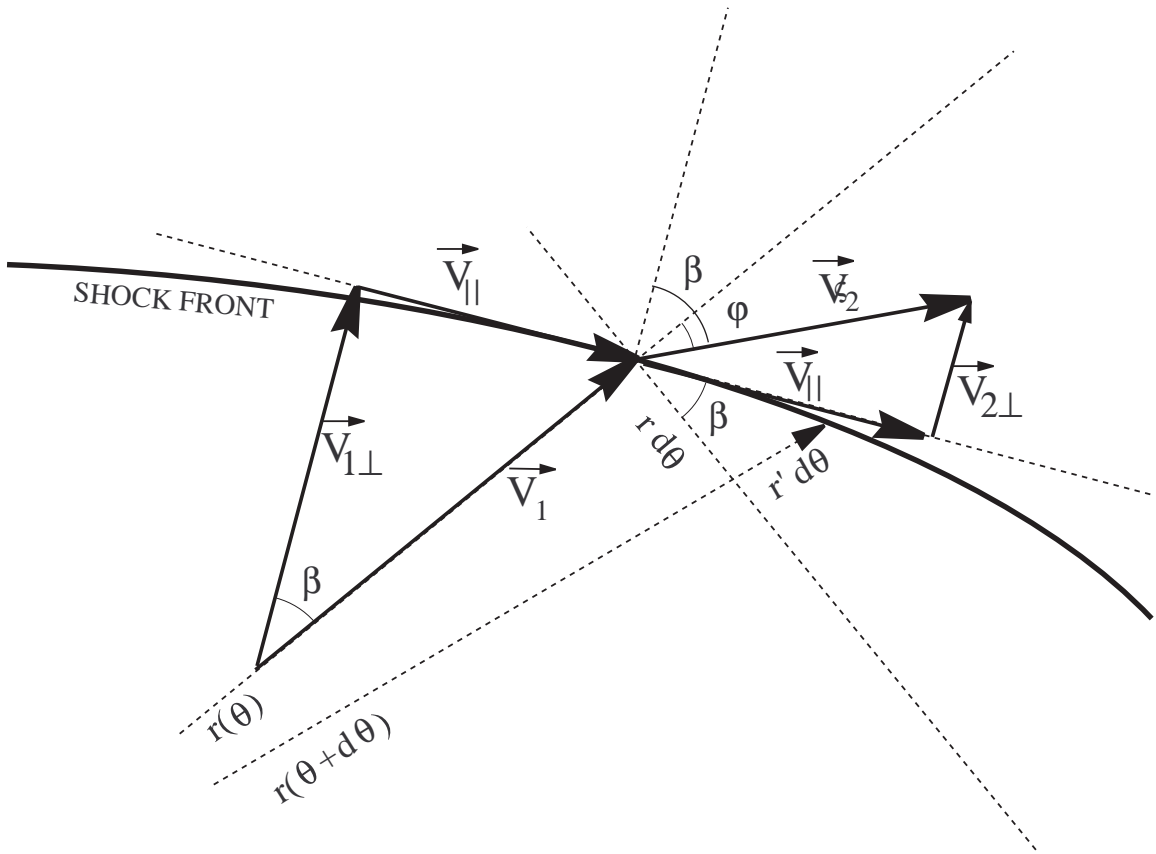}
\caption{The flow geometry near the shock front. \hfill}
\end{figure}

\begin{figure}
\centerline{\psfig{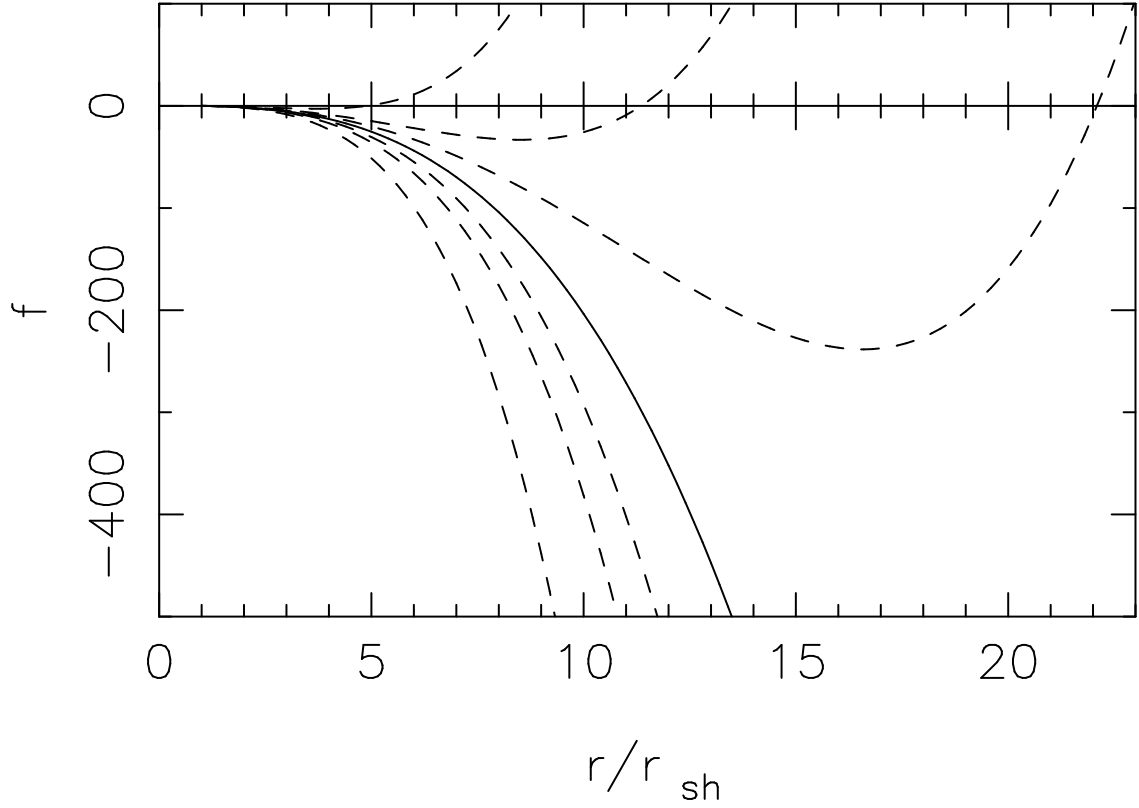}}
\caption{A family of solutions to Eq.~(A.30) Solid line is the
solution that corresponds to the absence of additional
perturbations in the postshock flow. \hfill}
\end{figure}

\begin{figure}
\centerline{\psfig{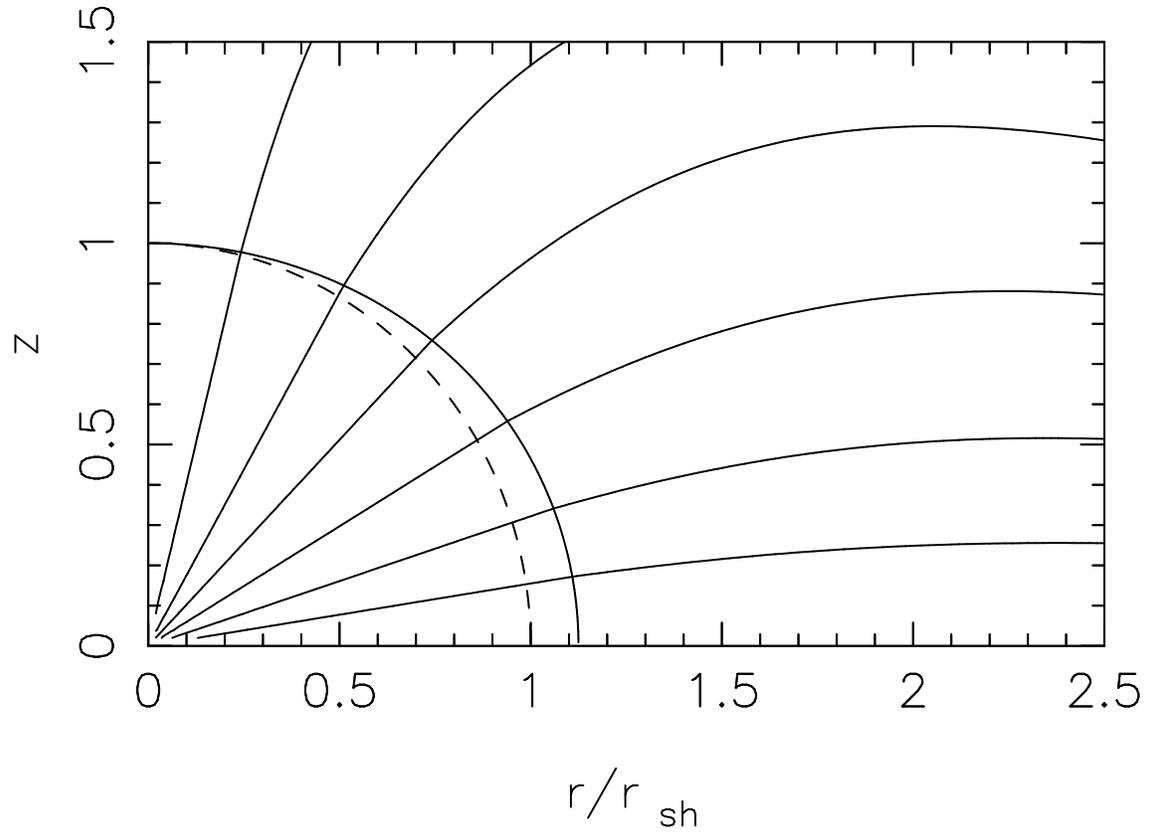}}
\caption{The distribution of streamlines for a flow with
$\alpha=0.25$. The shock location for isotropic (\emph{dashed
line}) and anisotropic (\emph{solid line}) flows. The bending of
the streamlines at the shock and the further bending of the flow
toward the equator result in the flow compression near the
equatorial plane. \hfill} \vspace*{-5pt}
\end{figure}
\end{document}